\documentclass[superscriptaddress,aps]{revtex4}
\usepackage{epsfig}
\usepackage{amsmath}
\usepackage{amssymb}

\begin{document}

\title{Clustering with shallow trees}

\author{M. Bailly-Bechet}
\affiliation{Universit\'e Lyon 1; CNRS UMR 5558; Laboratoire de Biom\'etrie et Biologie \'Evolutive; Villeurbanne, France}

\author{S. Bradde}
\affiliation{SISSA, via Beirut 2/4, Trieste, Italy and INFN Sezione di Trieste, Italy}

\author{A. Braunstein}
\affiliation{Politecnico di Torino, C.so Duca degli Abbruzzi 24, Torino, Italy}

\author{A. Flaxman}
\affiliation{IHME, University of Washington, Seattle, Washington, USA}

\author{L. Foini}
\affiliation{SISSA, via Beirut 2/4, Trieste, Italy and INFN Sezione di Trieste, Italy}

\author{R. Zecchina}
\affiliation{Politecnico di Torino, C.so Duca degli Abbruzzi 24, Torino, Italy}

\begin{abstract}

We propose a new method for hierarchical clustering based on the
optimisation of a cost function over trees of limited depth, and we
derive a message--passing method that allows to solve it efficiently.
The method and algorithm can be interpreted as a natural  interpolation
between two well-known approaches, namely single linkage and the recently presented Affinity Propagation.
We analyze  with this general scheme three  biological/medical structured datasets
(human population based on genetic information, proteins based on sequences and verbal autopsies)
and show that the interpolation technique provides  new insight.
\end{abstract}

\maketitle

\section{Introduction}

A standard approach to data clustering, that we will also follow here,
involves defining a distance measure between objects, called
dissimilarity. In this context, generally speaking data clustering 
deals with the problem
of classifying objects so that objects within the same class or
cluster are more similar than objects belonging to different
classes. The choice of both the measure of similarity and the
clustering algorithms are crucial in the sense that they define an
underlying model for the cluster structure.  In this work we discuss
two somewhat opposite clustering strategies, and show how they nicely fit as
limit cases of a more general scheme that we propose.

Two well-known general approaches that are extensively employed are
partitioning methods and hierarchical clustering methods \cite{Flynn}.
Partitioning methods are based on the choice of a given number of
\emph{centroids} -- \emph{i.e.} reference elements -- to which the
other elements have to be compared.  In this sense the problem reduces
to finding a set of centroids that minimises the cumulative distance
to points on the dataset. Two of the most used partitioning algorithms
are $K$-means (KM) and Affinity Propagation (AP)\cite{Leone,Frey2007}.
Behind these methods, there is the assumption of spherical
distribution of data: clusters are forced to be loosely of spherical shape,
with respect to the dissimilarity metric. These techniques give good
results normally only when the structure underlying the data fits this
hypothesis. Nevertheless, with Soft Affinity Propagation \cite{Leone}
the hard spherical constraint is relaxed, allowing for cluster
structures including deviation from the regular shape. This method
however recovers partially information on hierarchical organisation.  
On the other hand, Hierarchical Clustering methods such as single linkage
(SL) \cite{Michael}, starts by defining a cluster for each element of
the system and then proceeds by repeatedly merging the two closest
clusters into one. This procedure provides a hierarchic sequence of
clusters.

Recently an algorithm to efficiently approximate optimum spanning
trees with a maximum depth $D$ was presented in
\cite{BaBoBrChRaZe2008}.  We show here how this algorithm may be used
to cluster data, in a method that can be understood as a
generalisation of both (or rather an interpolation between) the AP and
SL algorithms. Indeed in the $D=2$ and $D=n$ limits - where $n$ is the
number of object to cluster - one recovers respectively AP and SL
methods.  As a proof of concept, we apply the new approach to a
collection of biological and medical  clustering problems on which
intermediate values of $D$ provide new interesting results.  
In the next section, we define an objective function for clustering based on the cost of
certain trees over the similarity matrix, and we devise a
message-passing strategy to optimise the objective function.  The
following section is devoted to recovering two known algorithms, AP
and SL, which are shown to be special cases for
appropriately selected values of the external parameters $D$.
Finally, in the last section we use the algorithm on three
biological/medical data clustering problems for which external
information can be used to validate the algorithmic
performance. First, we cluster human individuals from several
geographical origins using their genetic differences, then we tackle
the problem of clustering homologous proteins based only on their
amino acid sequences. Finally we consider a clustering problem arising
in the analysis causes-of-death in regions where vital registration
systems are not available.

\section{A Common Framework}

\begin{figure}
\epsfig{figure=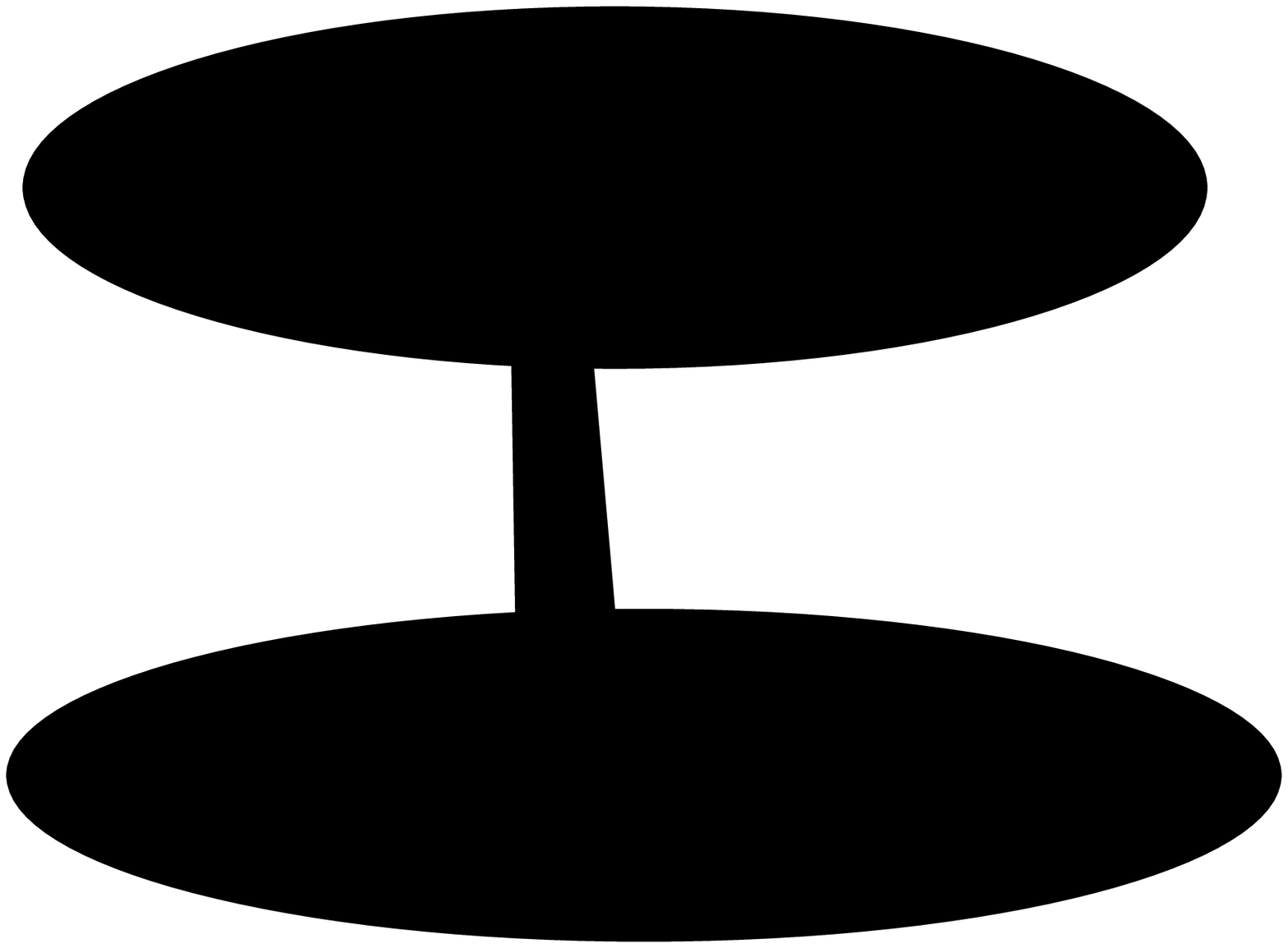,width=0.24\columnwidth}
\epsfig{figure=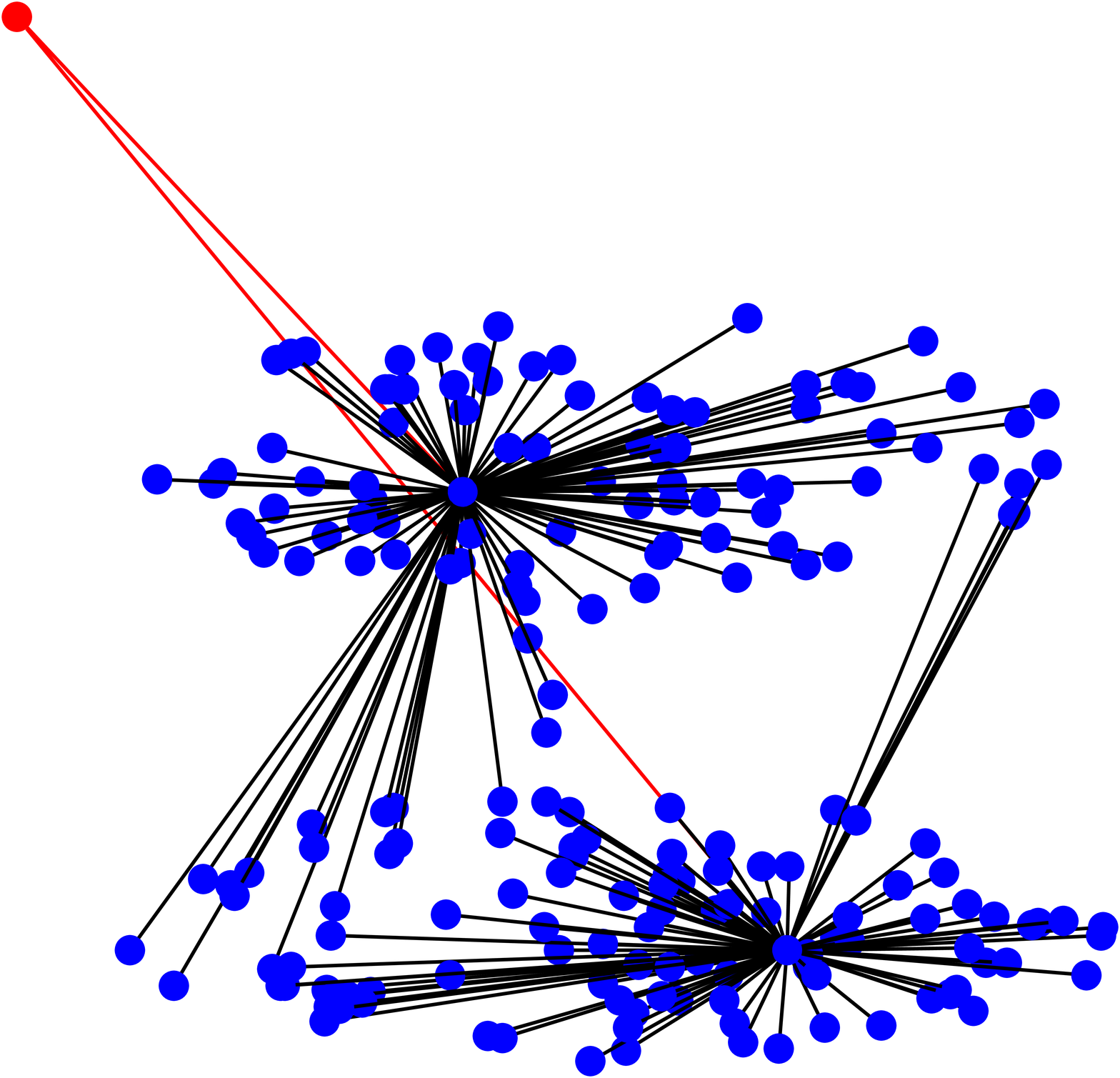,width=0.24\columnwidth}
\epsfig{figure=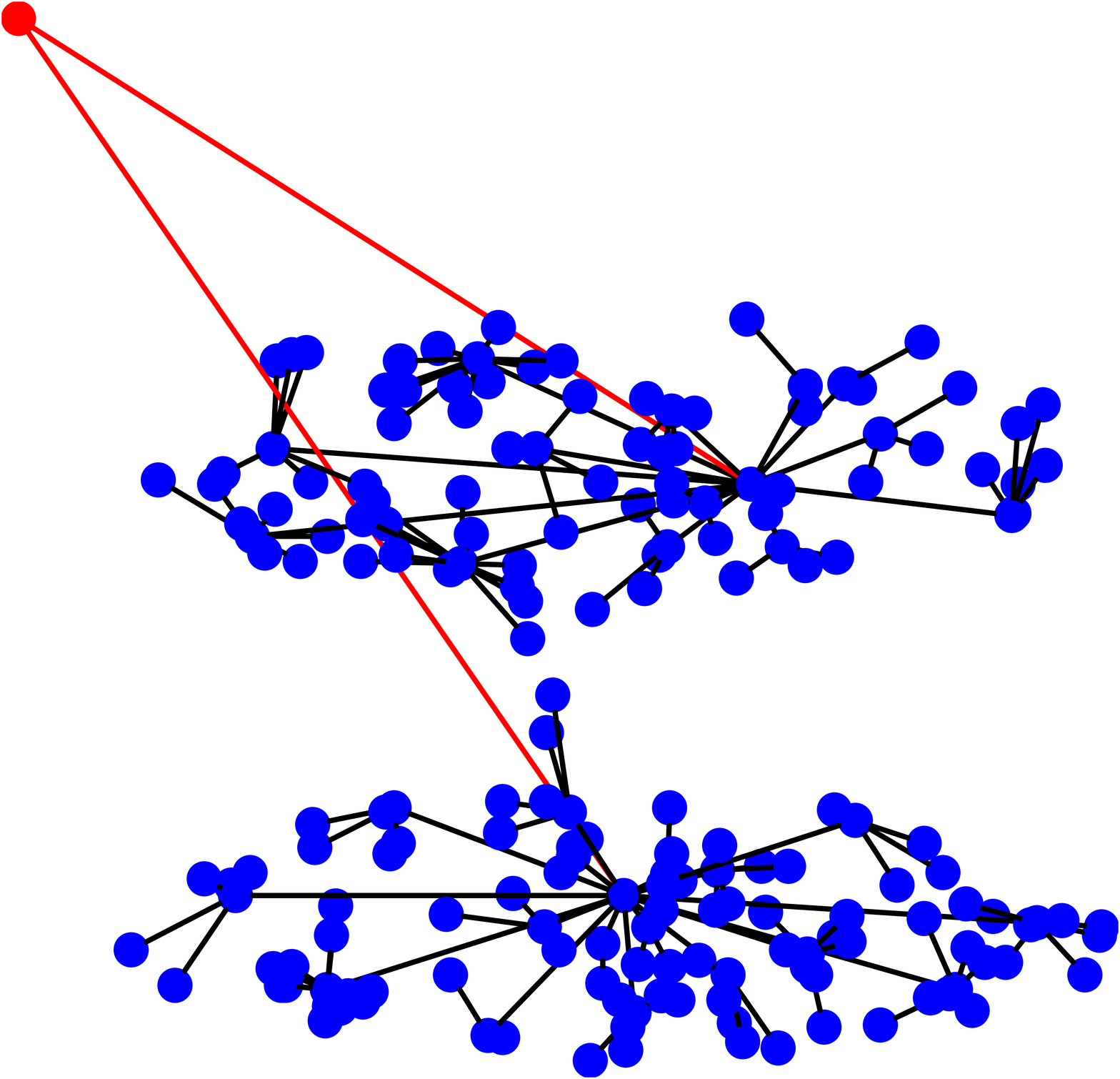,width=0.24\columnwidth}
\epsfig{figure=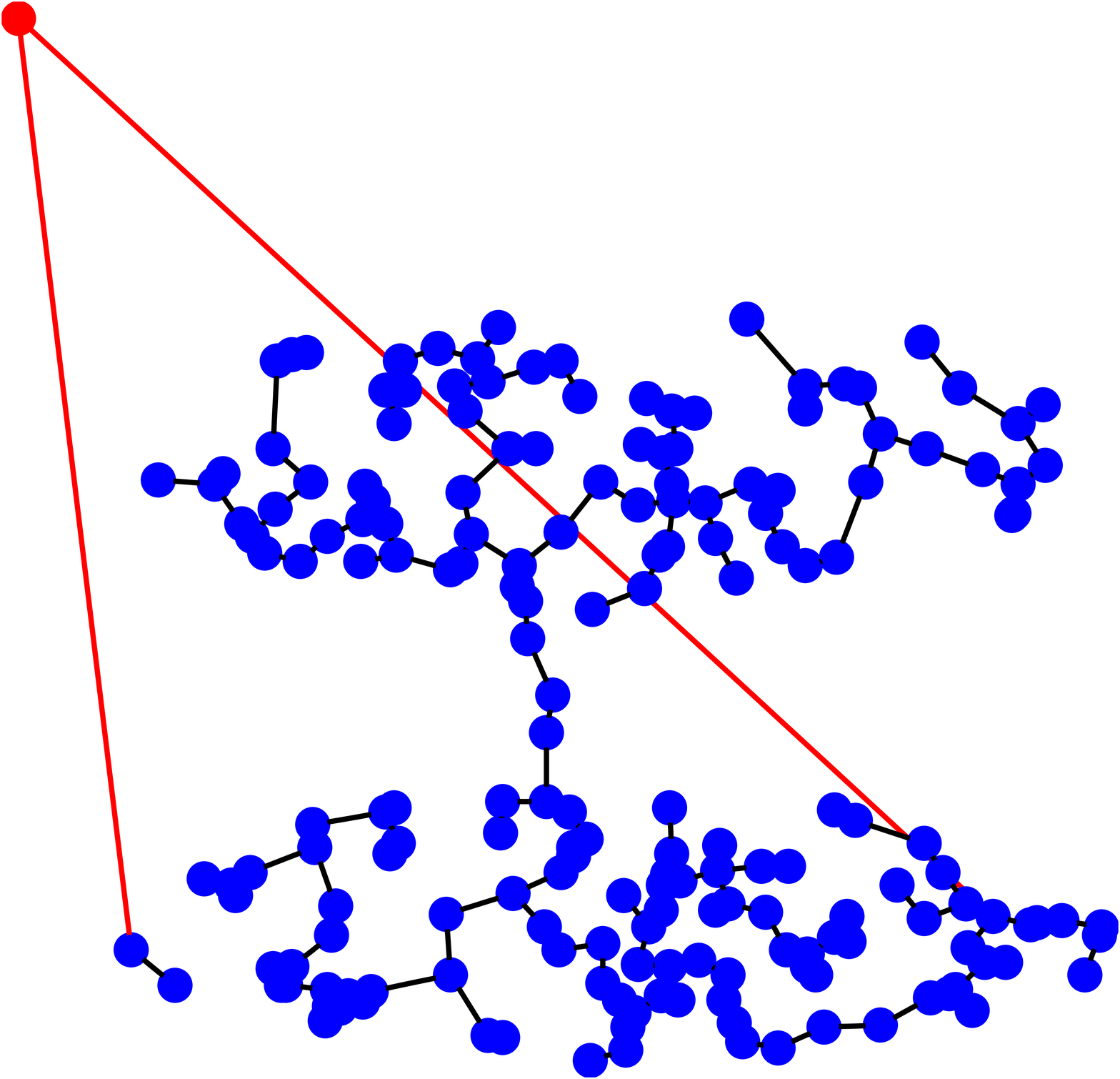,width=0.24\columnwidth}
\caption{Clustering an artificial 2D image. The black image on the
  left was randomly sampled and the euclidean distance was used as a
  measure of disimilarity between nodes. Clustering by $D$-MST was then attempted on
  the resulting graph. One external root vertex $v^*$ (red point) was added, with
  distance $\lambda$ to every other points. The ouput of the algorithm consists in 
  a minimum weight rooted spanning tree of depth $D$ pointed out by bold
  links.  The last three figures concern the resulting clustering for different 
  choices of the depth limit $D=2,4,>n$
  respectively. Different clusters with a complex internal structure
  can be recovered after removing the red node $v^*$. In the case of
  AP $D=2$ (second figure) the spherical clusters do not fit the ellipsoidal 
  shape of the original figure while for $4$-MST (third figure) the
  structure of two ellipses can be recovered.  The fourth and 
  last figure corresponds
  to SL ($D>n$): in this case nodes are split into two arbitrary
  components disregarding the original shape.}
\label{fig:spannfor}
\end{figure}

Let's start with some definitions. Given $n$ datapoints, we introduce the
similarity matrix between pairs $s_{i,j}$, where $i,j\in
[1,\ldots,n]$. This interaction could be represented as a fully
connected weighted graph $G(n,s)$ where $s$ is the weight associated
to each edge. This matrix contitutes the only data input for the clustering methods 
discussed in this manuscript. We refer in the following to the neighbourhood of node
$i$ with the symbol $\partial i$, denoting the ensemble of all nearest
neighbours of $i$.  By adding to the graph $G$ one artificial node
$v^*$, called {\em root}, whose similarity to all other nodes $i\in G$
is a constant parameter $\lambda$, we obtain a new graph
$G^*(n+1,s^*)$ where $s^*$ is a
$\left(n+1\right)\times\left(n+1\right)$ matrix with one added row and
column of constant value to the matrix $s$ (see Figure \ref{fig:spannfor}).

We will employ the following general scheme for clustering based on
trees. Given any tree $T$ that spans all the nodes in the graph
$G^*(n+1,s^*)$, consider the (possibly disconnected) subgraph
resulting of removing the root $v^*$ and all its links. We will define the
output of the clustering scheme as the family of vertex sets of the
connected components of this subgraph. That is, each cluster will 
be formed by a connected component of the pruned $T\setminus v^*$. 
In the following, we will concentrate on how to produce trees
associated to $G^*$.

The algorithm described in \cite{BaBoBrChRaZe2008} was devised to find 
a tree of minimum weight with a depth bounded by $D$ from a selected root to a
set of terminal nodes. In the clustering framework, all nodes are
terminals and must be reached by the tree.  As a tree has 
exactly $n-1$ links, for values of $D$ greater or equal than $n$ the
problem becomes the familiar (unconstrained) minimum spanning tree
problem. In the rest of this section we will describe the $D$-MST
message passing algorithm of \cite{BaBoBrChRaZe2008} for Steiner trees
in the simplified context of (bounded depth) spanning trees.

To each node of the graph we associate two variables $\pi_i$, and
$d_i$ where $\pi_i\in \partial i$ could be interpreted as a pointer
from $i$ to one of the neighbouring nodes $j\in \partial i$.
Meanwhile $d_i\in [0,\ldots, D]$ is thought as a discrete distance
between the node $i$ and the root $v^*$ along the tree. Necessarily
only the root has zero distance $d_{v^*}=0$, while for all other nodes
$d_i\in [1,\ldots,D]$.  In order to ensure global connectivity of the
$D$-MST, these two variables must satisfy the following condition:
$\pi_i=j \Rightarrow d_i=d_j+1$. This means that if node $j$ is the
parent of node $i$, then the depth of node $i$ must exceed the depth
of the node $j$ by precisely one.  This condition avoids the presence
of loops and forces the graph to be connected, assigning non-null
weight only to configurations corresponding to trees. The energy
function thus reads
\begin{equation}\label{eq:steinen}
  E(\{\pi_i,d_i\}_{i=1}^N)= \sum_i s_{i,\pi_i}-\sum_{i,j\in \partial i} \left( h_{ij}(\pi_i,\pi_j,d_i,d_j)+h_{ji}(\pi_j,\pi_i,d_j,d_i) \right),
\end{equation}
where $h_{ij}$ is defined as
\begin{equation}\label{eq:const}
  h_{ij}=\left\{\begin{array}{ll}
  0 & \{\pi_i=j \Rightarrow d_i=d_j+1\}\\\
  -\infty & \mbox{ else }\\
  \end{array}\right.
\end{equation}
In this way only configurations corresponding to a tree are taken into account with
the usual Boltzmann weight factor $e^{-\beta s_{i,\pi_i}}$ where the
external parameter $\beta$ fixes the value of energy level.  Thus the
partition function is
\begin{equation}
  Z(\beta)=\sum_{\{\pi_i,d_i\}} e^{ -\beta
    E(\{\pi_i,d_i\})}=\sum_{\{\pi_i,d_i\}}\prod_{i}e^{-\beta
    s_{i,\pi_i}}\times \prod_{i j\in \partial i} f_{ij},
\end{equation} 
where we have introduce an indicator function
$f_{ij}=g_{ij}g_{ji}$. Each term
$g_{ij}=1-\delta_{\pi_i,j}\left(1-\delta_{d_j,d_i-1}\right)$ is
equivalent to $e^ {h_{ij}}$ while $\delta_{ij}$ is the delta
function. In terms of these quantities $f_{ij}$ it is possible to derive
 the cavity equations, i.e. the following set of coupled equations for the cavity marginal
probability $P_{j\to i}(d_j,\pi_j)$ of each site $j \in [1,\ldots,n]$
after removing one of the nearest neighbours $i\in \partial j$:
\begin{eqnarray}\label{eq:cavity1}
&&P_{j\to i}(d_j,\pi_j)\propto e^{-\beta s_{i,\pi_i}}\prod_{k\in
    \partial j/i} Q_{k\to j}(d_j,\pi_j)\\
\label{eq:cavity2}
&&Q_{k\to j}(d_j,\pi_j)\propto \sum_{d_k \pi_k}P_{k\to j}(d_k,\pi_k)
f_{jk}(d_j,\pi_j,d_k,\pi_k)\,.
\end{eqnarray}
These equations are solved iteratively and in graphs with no cycles
they are guaranteed to converge to a fixed point that is the optimal
solution. In terms of cavity probability we are able to compute marginal and
joint probability distribution using the following relations
\begin{eqnarray}
&&P_j(d_j,\pi_j)\propto \prod_{k\in \partial j}Q_{k\to j}(d_j,\pi_j)\\
&&P_{ij}(d_i,\pi_i,d_j,\pi_j) \propto P_{i\to j}(d_i,\pi_i)P_{j\to i}(d_j,\pi_j) f_{ij}(d_i,\pi_i,d_j,\pi_j)\,.
\end{eqnarray}

For general graphs convergence can be forced 
by introducing a "reinforcement" perturbation term as in
\cite{BaBoBrChRaZe2008,braunstein2006learning}.  This leads to a new set of perturbed coupled 
equations that show good convergence properties.
The $\beta\to\infty$ limit is taken by considering the change of
variable $\psi_{j\to i}(d_j,\pi_j)=\beta^{-1} \log P_{j\to
  i}(d_j,\pi_j)$ and $\phi_{j\to i}(d_j,\pi_j)=\beta^{-1} \log Q_{j\to
  i}(d_j,\pi_j)$ then the relations \ref{eq:cavity1} and
\ref{eq:cavity2} reduce to
\begin{eqnarray}\label{eq:maxsum}
  &&\psi_{j\to i}(d_j,\pi_j)= -s_{i,\pi_i}+\sum_{k\in \partial j /i} \phi_{k\to j}(d_j,\pi_j)\\
\label{eq:maxsum1}
    &&\phi_{k\to j}(d_j,\pi_j)= \max_{d_k \pi_k: f_{kj}\neq 0}\psi_{k\to j}(d_k,\pi_k)\,.
\end{eqnarray}
These equations are in the Max-Sum form and equalities hold up
to some additive constant. In terms of these quantities marginals are
given by $\psi_j(d_i,\pi_j)=-c_{j\pi_j}+\sum_k \phi_{k\to j}
(d_j,\pi_j)$ and the optimum tree is the one obtained by
$\mbox{argmax}\; \psi_j$.  If we introduce the variables $A^d_{k\to
  j}=\max_{\pi_k\neq j}\psi_{k\to j}(d,\pi_k)$, $C^d_{k\to
  j}=\psi_{k\to j}(d,j)$ and $E^d_{k\to j}=\max(C^d_{k\to j},A^d_{k\to
  j})$ it is enough to compute all the messages $\phi_{k\to
  j}(d_j,\pi_j)=A^{d_j-1}_{k\to j},E^{d_j}_{k\to j}$ for $\pi_j=k$ and
$\pi_j\neq k$ respectively. Using equations \ref{eq:maxsum} and
\ref{eq:maxsum1} we obtain the following set of equations:
\begin{eqnarray}
&&A^d_{j\to i}(t+1)=\sum_{k\in N(j)/i}E^d_{k\to j}(t)+ \max_{k\in N(j)/ i}\left(A^{d-1}_{k\to j}(t) -E^d_{k\to j}(t)-s_{j,k}\right)\\
&&C^d_{j\to i}(t+1)= -s_{j,i}+\sum_{k\in N(j)/i}E^d_{k\to j}(t)\\
&&E^d_{j\to i}(t+1)=\max\left(C^d_{j\to i}(t+1),A^d_{j\to i}(t+1)\right)\,.
\end{eqnarray}

It has been demonstrated \cite{spanningtree} that a fixed point of
these equations with depth $D>n$ is an optimal spanning tree.  In the
following two subsections, we show how to recover the SL and AP
algorithms. On one hand, by computing the (unbounded depth) spanning
tree on the enlarged matrix and then considering the connected
components of its restriction to the set of nodes removing $v^{*}$, we
recover the results obtained by SL. On the other hand we obtain AP by
computing the $D=2$ spanning tree rooted at $v^{*}$, defining the
self-affinity parameter as the weight to reach this root node.

\subsection{Single Linkage limit}

Single Linkage is one of the oldest and simplest clustering methods,
and there are many possible descriptions of it. One of them is the
following: order all pairs according to distances, and erase as many
of the pairs with largest distance so that the number of resulting
connected components is exactly $k$. Define clusters as the resulting
connected components.

An alternative method consists in removing initially all
\emph{useless} pairs (i.e. pairs that would not change the set of
components when removed in the above procedure).  This reduces to the
following algorithm: given the distance matrix $s$, compute the
minimum spanning tree on the complete graph with weights given by
$s$. From the spanning tree remove the $k-1$ links with largest
weight. Clusters are given by the resulting connected components. In
many cases there is no \emph{a priori } desired number of clusters $k$
and an alternative way of choosing $k$ is to use a continuous
parameter $\lambda$ to erase all weights larger than $\lambda$.

The $D$-MST problem for $D>n$ identifies the minimum spanning tree
connecting all $n+1$ nodes (including the root $v^*$). This means each node 
$i$ will point to one other node $\pi_i=j\neq v^*$
if its weight satisfies the condition $\min_j s_{i,j}< s_{i,v^*}$,
otherwise it would be cheaper to connect it to the root (introducing
one more cluster). We will make this description more precise. For simplicity, 
let's assume no edge in $G(n,s)$ has weight exactly equal to $\lambda$.

The Kruskal algorithm \cite{Kruskal} is a classical algorithm to compute a 
minimum spanning tree. It works by iteratively creating a forest as follows: start 
with a subgraph all nodes and no edges. The scan the list of edges ordered 
by increasing weight, and add the edge to the forest if it connects two 
different components (i.e. if it does not close a loop). At the end of the 
procedure, it is easy to prove that the forest has only one connected
component that forms a minimum spanning tree.
It is also easy to see that the edges added when applying the Kruskal algorithm 
to $G(n,s)$ up to the point when the weight reaches $\lambda$ 
are also admitted on the Kruskal algorithm for $G(n+1,s^*)$. After that point, 
the two procedures diverge because on $G(n,s)$ the remaining added edges have weight larger 
than $\lambda$ while on $G(n+1,s^*)$ all remaining added edges have weight exactly $\lambda$.
Summarizing, the MST on $G(n+1,s^*)$ is a MST on $G(n,s)$ on which all edges with weight 
greater than $\lambda$ have been replaced by edges connecting with $v^*$.

\subsection{Affinity propagation limit}

Affinity Propagation is a method that was recently proposed in
\cite{Frey2007}, based on the choice of a number of ``exemplar''
data-points. Starting with a similarity matrix $s$, choose a set of exemplar
datapoints $X\subset V$ and an assignment $\phi:V\mapsto X$ such that: $\phi(x)=x$ 
if $x\in X$ and the sum of the distances between datapoints and the exemplars they map
to is minimised.  It is essentially based on iteratively passing two
types of messages between elements, representing \emph{responsibility}
and \emph{availability}.  The first, $r_{i \to j}$, measures how much
an element $i$ would prefer to choose the target $j$ as its
exemplar. The second $a_{i\to j}$ tells the preference for $i$ to be
chosen as an exemplar by datapoint $j$. This procedure 
is an efficient implementation of the Max-Sum algorithm that improves
the naive exponential time complexity to $O(n^2)$.  The
self-affinity parameter, namely $s_{i, i}$, is chosen as the
dissimilarity of an exemplar with himself, and \emph{in fine}
regulates the number of groups in the clustering procedure, by
allowing more or less points to link with ``dissimilar''
exemplars.\\ Given a similarity matrix $s$ for $n$ nodes, we want to
identify the \emph{exemplars}, that is, to find a valid configuration
$\overline{\pi}=\{\pi_1,\ldots,\pi_n\}$ such that $\pi: [1,\ldots,n]
\mapsto [1,\ldots,n]$ so as to minimise the function
\begin{equation} 
  E(\overline{\pi})= -\sum_{i=1}^n s_{i,\pi_i}-\sum_i \delta_i(\overline{\pi})\, ,
\end{equation}
where the constraint reads
\begin{equation}
  \delta_i(\overline{\pi})=\left \{\begin{array}{ll}
  -\infty &   \pi_i\neq i \;\cap\; \exists\; j :\; \pi_j=i\\
  0 & \text{else}
  \end{array}\right.
\end{equation}
These equations take into account the only possible configurations,
where node $i$ either is an exemplar, meaning $\pi_i=i$, or it is not
chosen as an exemplar by any other node $j$. The energy function thus
reads
\begin{equation}\label{eq:affen}
  E(\overline{\pi})=\left \{\begin{array}{ll}
  -\sum_i s_{i,\pi_i} &   \forall\; i\;\{\; \pi_i= i \;\cup\; \forall j\; \pi_j\neq i\;\}\\
  \infty & \mbox{ else }
  \end{array}\right.
\end{equation}
The cavity equations are computed starting from this definition and
after some algebra they reduce to the following update conditions for 
responsibility and availability \cite{Frey2007}:
\begin{eqnarray}\label{affinity}
 && r_{i\to k}^{t+1}= s_{i, k}-\max_{k'\neq k}\left( a_{k'\to i}^t+s_{k', i}\right)\\
  && a_{k\to i}^{t+1}=\min\left(0,r_{k\to k}+\sum_{i' \neq k\; i} \max\left(0,r_{i'\to k}^t\right)\right)\,.
\end{eqnarray}

In order to prove the equivalence between the two algorithms,
i.e. $D$-MST for $D=2$ and AP, we show in the following how the two 
employ an identical decomposition of the same energy function thus resulting
necessarily to the same max sum equations. In the $2$-MST equations,
we are partitioning all nodes into three groups: the first one is
only the root whose distance $d=0$, the second one is composed of
nodes pointing at the root $d=1$ and the last one is made up of nodes
pointing to other nodes that have distance $d=2$ from the root.  The
following relations between $d_i$ and $\pi_i$ makes this condition
explicit:
\begin{equation}
  d_i=\left\{\begin{array}{cc}
  1 & \Leftrightarrow \pi_i=v^*\\
  2 & \Leftrightarrow \pi_i\neq v^*\\
  \end{array}\right.
\end{equation}
It is clear that the distance variable $d_i$ is redundant because
the two kind of nodes are perfectly distinguished with just the
variable $\pi_i$.  Going a step further we could remove the external
root $v^*$ upon imposing the following condition for the pointers
$\pi_i= i \Leftrightarrow \pi_i=v^* \; \pi_i=j\neq i \Leftrightarrow
\pi_i\neq v^*$.  This can be understood by thinking at AP procedure:
since nodes at distance one from the root are the exemplars, they
might point to themselves, as defined in AP, and all the non-exemplars
are at distance $d=2$ so they might point to nodes at distance $d=1$. Using 
this translation, from Equation \ref{eq:const} it follows that
\begin{equation}\label{eq:const1}
\sum_{ij\in \partial i}h_{ij}+h_{ji}=\left\{\begin{array}{cc}
  0 & \forall\; i\; \{\pi_i=i\;  \cup \;\forall\; j\neq i\; \pi_j\neq i\}\\
  -\infty & \mbox{ else}
  \end{array}\right.
\end{equation} 
meaning that the constraints are equivalent 
$\sum_{ij\in \partial i} h_{ij}+h_{ji}=\sum_i \delta_i(\overline{\pi})$. 
Substituting (\ref{eq:const1}) into equation (\ref{eq:steinen})
we obtain that 
\begin{equation}
  E(\{\pi_i,d_i\}_{i=1}^n)=\left\{\begin{array}{ll}
  -\sum_{i}s_{i,\pi_i} & \forall \;i \{\pi_i=i\;  \cup \;\forall\; j\neq i\; \pi_j\neq i\}\\
  \infty & \mbox{ else }
  \end{array}\right.
\end{equation}
The identification of the self affinity parameter and the self 
similarity $s_{i,v^*}=\lambda = s_{i,i}$  allows us to prove the equivalence 
between this formula and the AP energy given in equation (\ref{eq:affen})
 as desired. 

\begin{figure}
A\epsfig{file=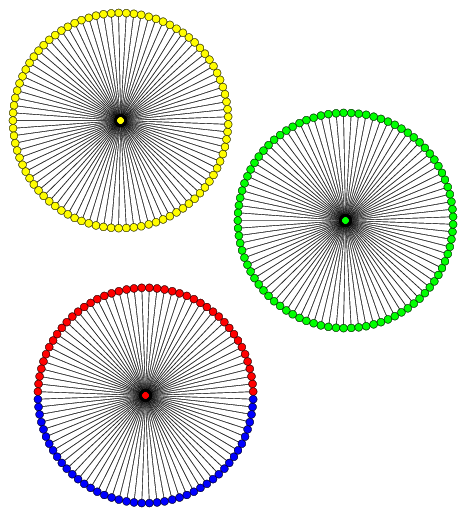,width=0.15\columnwidth}
\hspace{1cm}
B\epsfig{file=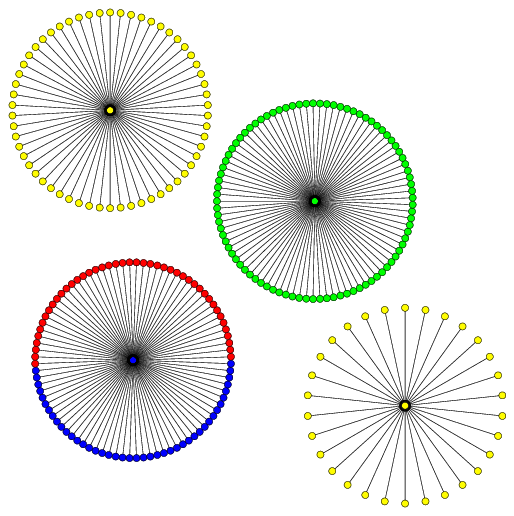,width=0.15\columnwidth}
\hspace{1cm}
C\epsfig{file=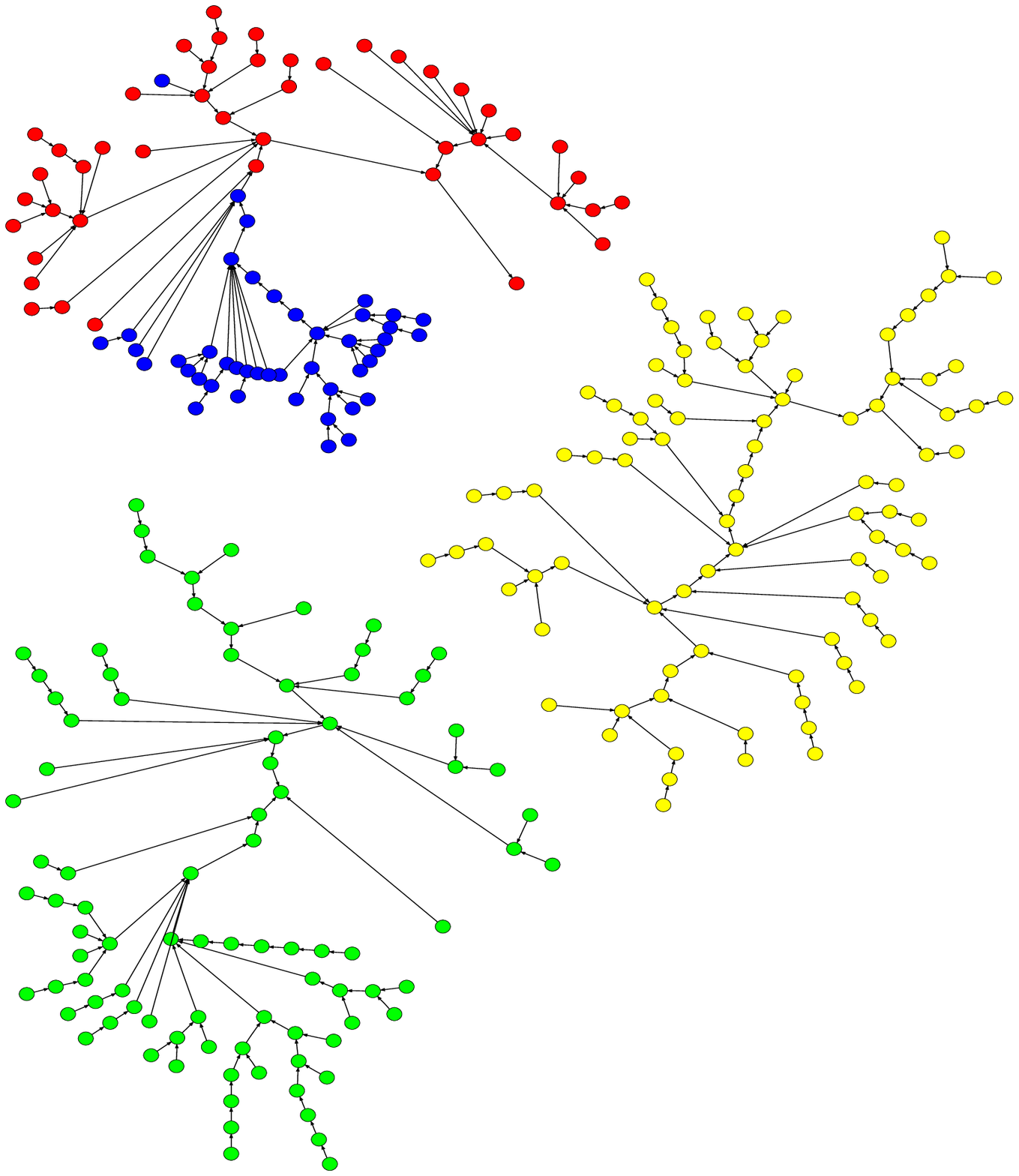,width=0.20\columnwidth}
D\epsfig{file=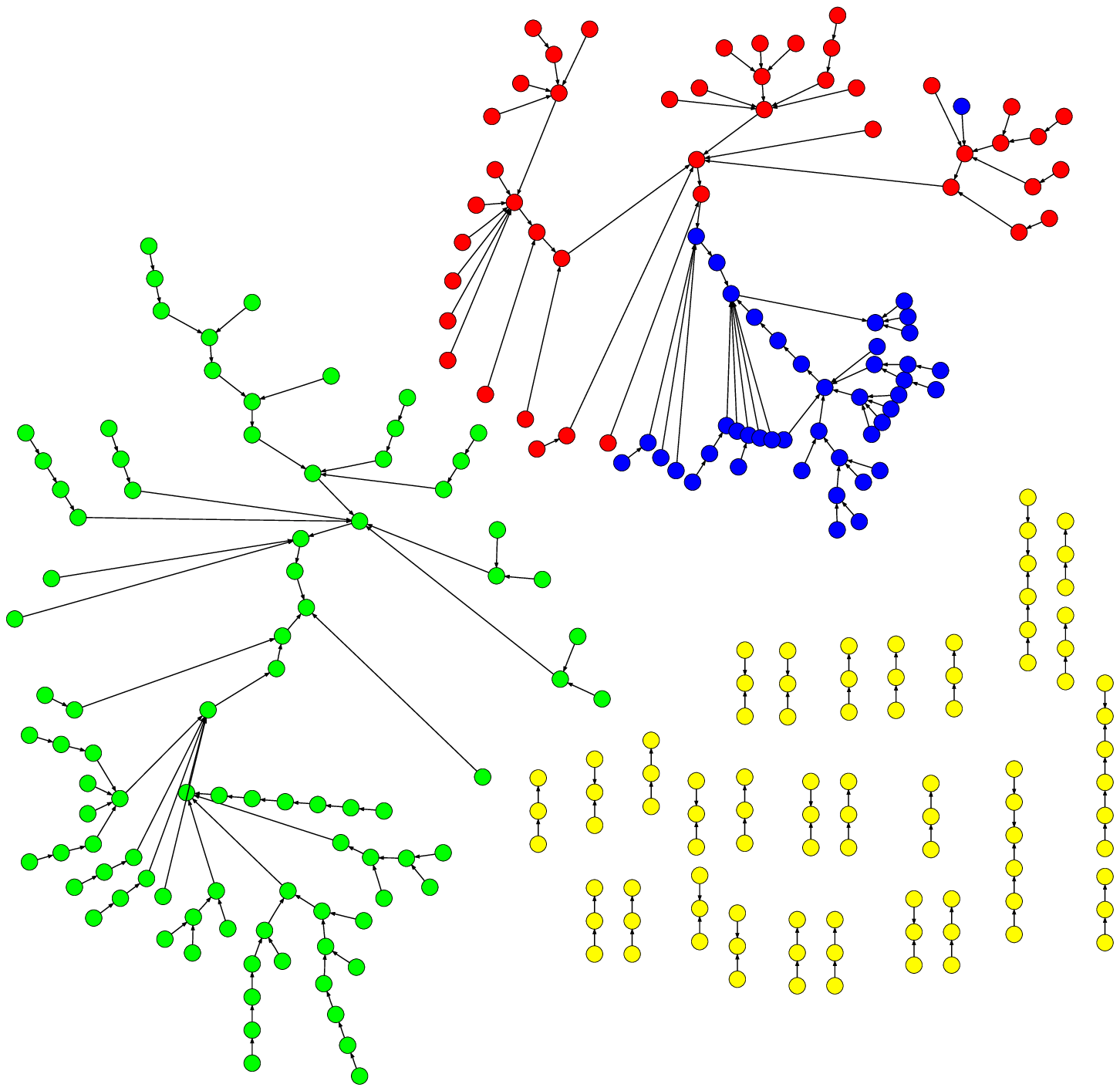,width=0.20\columnwidth}
\caption{In this figure we compare the results of Single Linkage and 
Affinity Propagation techniques on a SNP distance dataset. 
The dataset is composed of 269 individuals divided into four populations: 
CHB (red), CEU (green), YRI (yellow) and JPT (blue). 
The panels A and B are AP results while panels C and D
show clusters obtained with SL. As $\lambda$ increases, both algorithms fail 
to divide Chinese from Japanese populations (panels A, C) before 
splitting the Nigerian population (yellow).
}\label{apslclu}
\end{figure}

\section{Applications to biological data}

In the following sections we shall apply the new technique to different clustering problems
and give a preliminary comparison  to the two extreme limits of the interpolation, namely $D=2$ (AP) and $D=n$ (SL).

Clustering is a widely used method of analysis in biology, most notably in the
recents fields of transcriptomics \cite{Eisen1998}, proteomics and
genomics\cite{Barla2008}, where huge quantities of noisy data are
generated routinely. A clustering approach presents many advantages for this
type for data: it can use all pre-existing knowledge available to
choose group numbers and to assign elements to groups, it has good
properties of noise robustness\cite{Dougherty2002}, and it is
computationally more tractable than other statistical
techniques. In this section apply our
algorithm to structured biological data, in order to show that by
interpolating between two well-known clustering methods (SL and AP) 
it is possible to obtain new insight.

\subsection{Multilocus genotype clustering}
In this application we used the algorithm to classify individuals 
according to their original population using only information from 
their sequence SNPs as a distance measure\cite{aaa}.  A 
single-nucleotide polymorphism (SNP) is a DNA sequence variation
occurring when a single nucleotide (A, T, C, or G) in the genome (or
other shared sequence) differs between members of a species (or
between paired chromosomes in a diploid individual).  The dataset we
used is from the HapMap Project, an international project launched in
2002 with the aim of providing a public resource to accelerate medical
genetic research \cite{Hapmap}.  It consists of SNPs data from 269
individuals from four geographically diverse origins: 90 Utah
residents from North and West Europe (CEU), 90 Yoruba of Ibadan,
Nigeria (YRI), 45 Han Chinese of Beijing (CHB) and 44 Tokyo Japanese
(JPT). CEU and YRI samples are articulated in thirty families of three
people each, while CHB and JPT have no such structure.  For each
individual about 4 millions SNPs are given, allocated on different
chromosomes.  In the original dataset some SNPs were defined only in
subpopulations, thus we extracted those which were well-defined for
every sample in all populations and after this selection the number of
SNPs for each individual dropped to $1.8$ million.  We defined the
distance between samples by the number of different alleles on the
same locus among individuals normalised by the total number of
counts. The $269\times269$ matrix of distance $S$ was defined as
follows:
\begin{equation}
 s_{i,j} = \frac{1}{2N} \sum_{n=1}^{N} d_{ij}(n),
\end{equation} 
where $N$ is the number of valid SNPs loci and $d_{ij}(n)$ is the distance between the $n$-th genetic loci of individuals $i$ and $j$:
\begin{equation}
 d_{ij}(n) = 
\begin{cases}
0&\mbox{if $i$ and $j$ have two alleles in common at the $n$-th locus} \\
1&\mbox{if $i$ and $j$ share only one single allele in common} \\
2&\mbox{if $i$ and $j$ have no alleles in common}
\end{cases}
\end{equation} 
\begin{figure}
\begin{center}
\epsfig{file=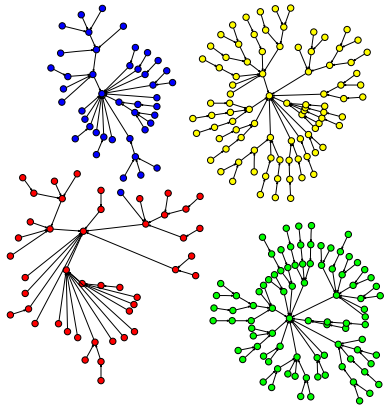,scale=0.7}
\end{center}
\caption{In this figure we report the clustering by $D$-MST with a fixed
  maximum depth $D=5$. The algorithm gives only one mis-classification.\label{stclu}}
\end{figure}

The resulting distance matrix was given as input to $D$-MST algorithm.
In figure \ref{stclu} we show the clusters found by the algorithm
using a maximum depth $D = 5$. Each individual is represented by a
number and coloured according to the population he belongs to: green
for YRI, yellow for CEU, blue for JPT and red for CHB. One can see
that the algorithm recognises the populations grouping the individuals
in four clusters. There is only one misclassified case, a JPT
individual placed in the CHB cluster.\\ Moreover, noticing that yellow
and green clusters have a more regular internal structure than the
other two, it is possible to consider them separately. Therefore, if
one applies the $D$-MST algorithm to this restricted subset of data,
all families consisting of three people can be immediately recovered,
and the tree subdivides in 60 families of 3 elements, without any
error (details not reported).

This dataset is particularly hard to classify, due to 
the complexity of the distances distribution. In fact the presence of 
families creates a sub-clustered structure inside the groups 
of YRI and CEU individuals. Secondly CHB and JPT people, 
even if  they belong to different populations, share in general smaller 
distances with respect to those subsisting among different families 
inside one of the other two clusters.
The $D$-MST algorithm overcomes this subtleties  with the possibility 
of developing a complex structure and allows the correct detection of 
the four populations  while other algorithms, such as AP, cannot adapt 
to this variability of the typical distance scale between groups in the dataset.
Indeed, the hard constraint in AP relies strongly on cluster-shape
regularity and forces clusters to appear as star of radius one: there
is only one central node, and all other nodes are directly connected
to it. Elongated or irregular multi-dimensional data might have more
than one simple cluster centre. In this case AP may force division of
single clusters into separate ones or may group together different
clusters, according to the input self-similarities.  Moreover, since
all data points in a cluster must point to the same exemplar, all
information about the internal structure, such as families grouping,
is lost. Thus, after running AP, CHB and JPT are grouped in a unique
cluster, and both CHB and JPT have the same exemplar, as shown on
figure \ref{apslclu}A. Going a step further and forcing the algorithm
to divide Chinese from Japanese we start to split the YRI population
(Fig. \ref{apslclu}B).\\ Hierarchical Clustering also fails on
this dataset, re\-co\-gni\-sing the same three clusters found by
Affinity Propagation at the 3-clusters level (Fig. \ref{apslclu}C) and
splitting yellow population in families before dividing blue from red
(see figure \ref{apslclu}D).  This makes sense relative to the typical
dissimilarities between individuals, but prevents grasping the
entire population structure.

\begin{figure}
\begin{center}
\epsfig{file=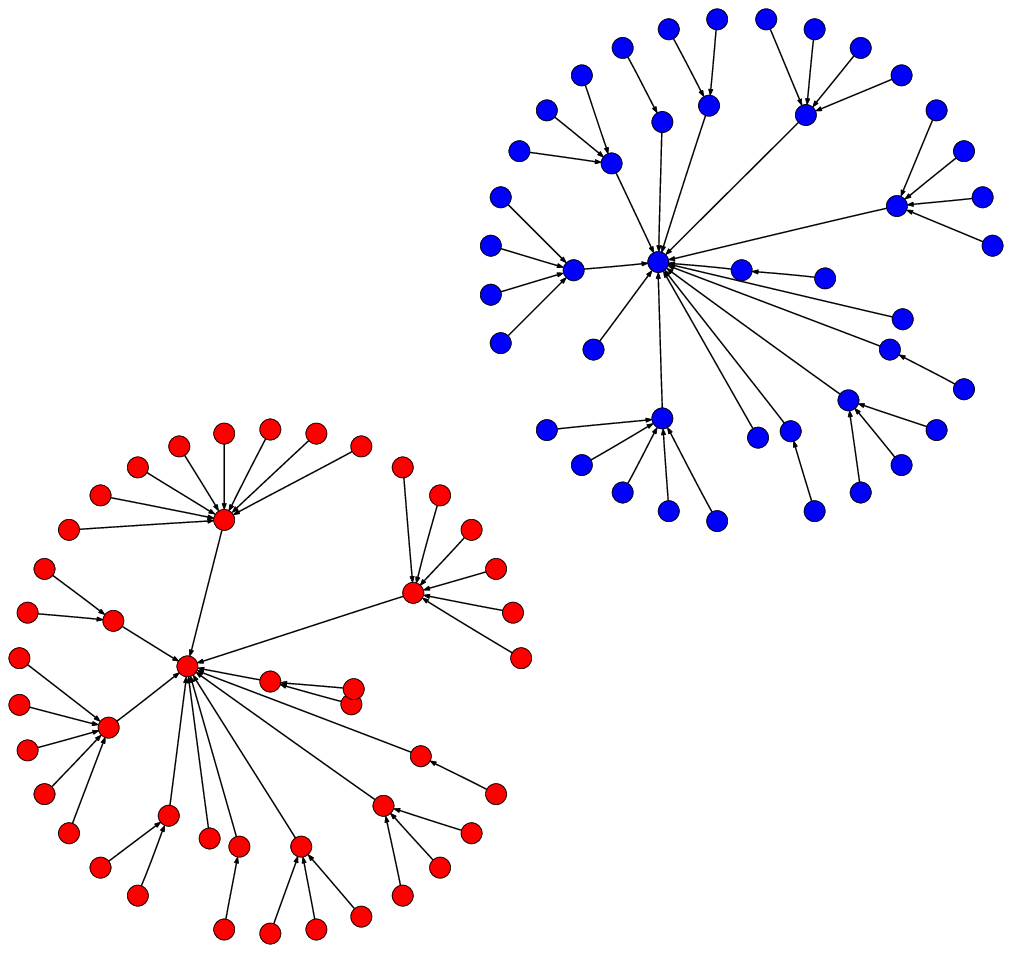,width=0.35\columnwidth}\hspace{1cm}
\epsfig{file=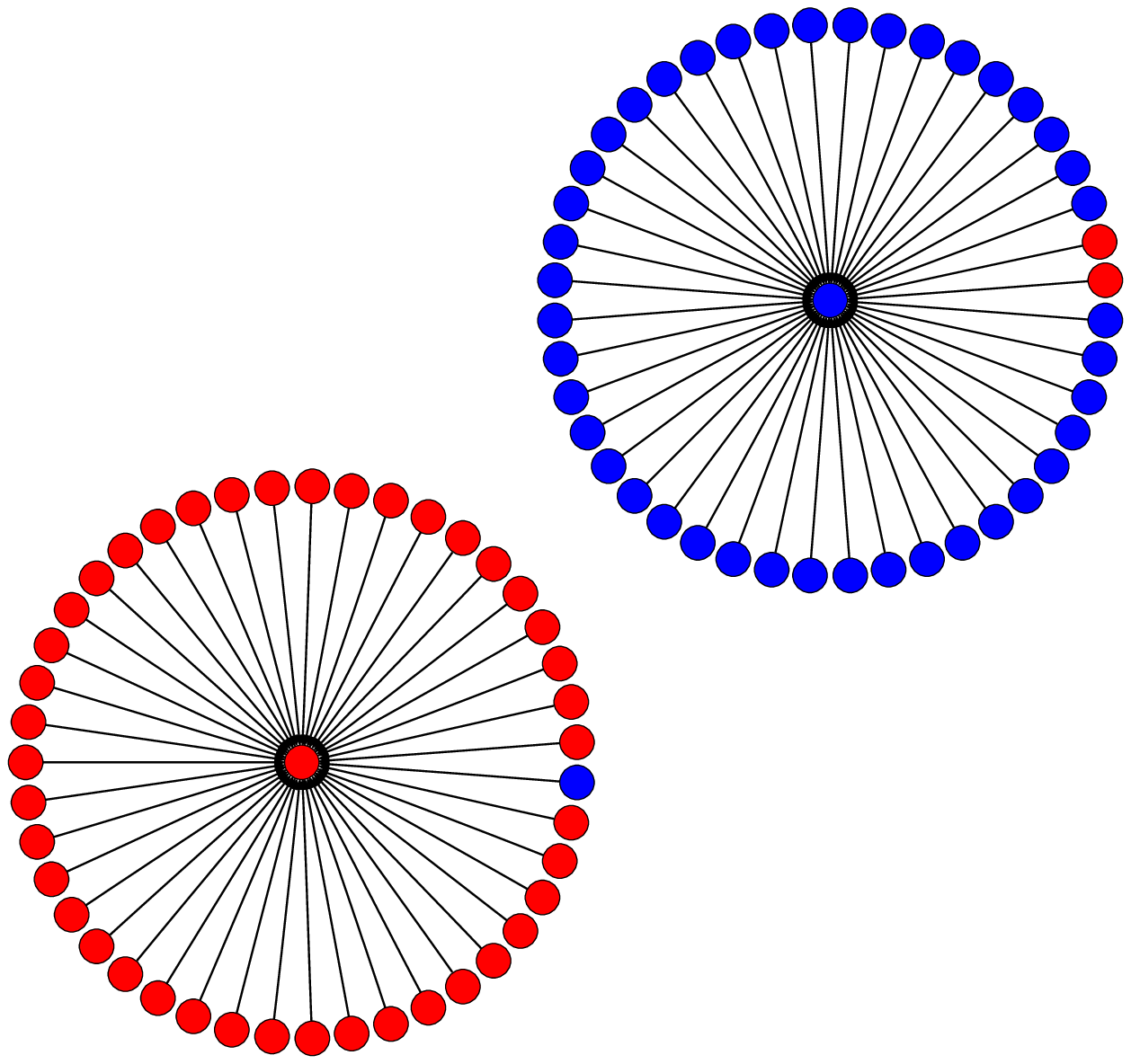,width=0.35\columnwidth}
\end{center}
\caption{We report clusters results obtained using
  the $D$-MST algorithm with $D=2$ (right) and $D=3$ (left), 
  considering only CHB and JPT. While both algorithms perform well 
  in this subset, the $3$-MST algorithm correctly 
  classifies all individuals.\label{s89}}
\end{figure}

After considering all four populations together, we applied the
$D$-MST algorithm only to the subset consisting of CHB and JPT
individuals, because these data appeared the hardest to cluster
correctly. The result is that the $D$-MST algorithm, with depth $D =
3$ succeeds in correctly detecting the two clusters without any
miss-classification as shown in the left part of figure \ref{s89}. Limiting the
analysis to this selected dataset, Affinity Propagation identifies two
different clusters, and is unable to divide CHB from JPT, still
committing three mis-classifications, as viewed in right panel of the
figure \ref{s89}.

The cluster structure found is controlled both by the maximum depth
$D$ and by $\lambda$, the two input parameters. In fact, in the
extreme case $D=1$ all samples in the dataset would
be forced to point to the root, representing a single cluster. The
next step is $D=2$ and corresponds to Affinity Propagation. This
allows to identify more than trivial clusters, as in the previous
case, but, as we said, still imposes a strong constraint, which, in
general, may be not representative of the effective data
distribution.   Increasing $D$ one can detect more structured
clusters, where different elements in the same group do not necessary
share a strong similarity with the same reference exemplar, as it has
to be with Affinity Propagation or $K$-means. On the other hand, the
possibility to detect an articulated shape gives some information
about the presence of eventual internal sub-structures notable to be
analysed separately, as in our case the partition of two groups into
families.

The parameter $\lambda$ also affects the result of the clustering, in
particular on the number of groups found. Assigning a large value to
this parameter amounts to pay a high cost for every node connected to
the root, and so to reduce the number of clusters; on the other side
decreasing $\lambda$ will create more clusters. In this first
application we used a value of $\lambda$ comparable with the typical
distance between elements, allowing us to detect the four clusters. In
this regime, one can expect a competition between the tendency of the
elements to connect to other nodes or to form new clusters with links
to the root, allowing the emergence of the underlying structure in the
data.

\subsection{Clustering of proteins datasets}

An important computational problem is grouping proteins into families
according to their sequence only. Biological evolution lets proteins
fall into so-called families of similar proteins - in term of
molecular function - thus imposing a natural classification. Similar
proteins often share the same three-dimensional folding structure,
active sites and binding domains, and therefore have very close
functions. They often - but not necessarily - have a common ancestor,
in evolutive terms. To predict the biological properties of a protein
based on the sequence information alone, one either needs to be able
to predict precisely its folded structure from its sequence
properties, or to assign it to a group of proteins sharing a known
common function. This second possibility stems almost exclusively from
properties conserved in through the evolutionary time, and is computationally much
more tractable than the first one. We want here to underline how our clustering method could be useful to
handle this task, in a similar way as the one we used in the first
application, by introducing a notion of distance between proteins
based only on their sequences.  The advantage of our algorithm
is its global approach: we do not take into account only distances
between a couple of proteins at a time, but we solve the clustering
problem of finding all families in a set of proteins in a
\emph{global} sense. This allows the algorithm to detect cases where
related proteins have low sequence identity.

To define similarities between proteins, we use the BLAST E-value as a
distance measure to assess whether a given alignment between two
different protein sequences constitutes evidence for homology. This
classical score is computed by comparing how strong an alignment is
with respect to what is expected by chance alone. This measure
accounts for the length of the proteins, as long proteins have more
chance to randomly share some subsequence.  In essence, if the E-value
is 0 the match is perfect while the more E-value is high the more the
average similarity of the two sequences is low and can be considered
as being of no evolutionary relevance.  We perform the calculation in
a all-by-all approach using the BLAST program, a sequence comparison
algorithm introduced by Altshul et al. \cite{Altshul}.\\ Using this
notion of distance between proteins we are able to define a matrix of
similarity $s$, in which each entry $s_{i,j}$ is associated to the
E-value between protein $i$ and $j$. The $D$-MST algorithm is then
able to find the directed tree between all the sets of nodes
minimising the same cost function as previously. The clusters we found
are compared to those computed by other clustering methods in the
literature, and to the ``real'' families of function that have been
identified experimentally.

\begin{figure}
\begin{center}
\epsfig{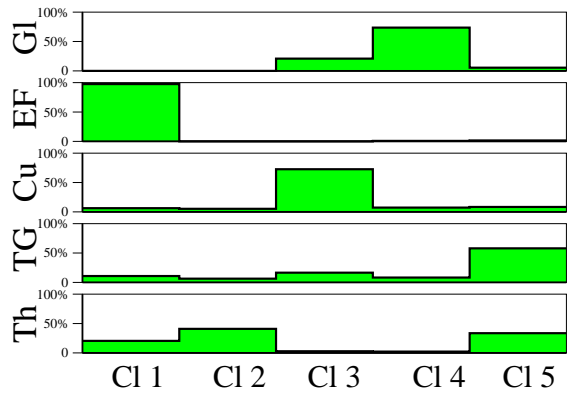}
\end{center}
\caption{We show the results of clustering proteins of the $5$ subfamilies 
Globin-like (Gl), EFhand (EF), Cupredoxin (Cu), Trans-Glycosidases (TG), 
Thioredoxin-like (Th) using $4$-MST with parameter $\lambda$=1.45. 
We see that most of the proteins of the first three families 
(Gl, EF and Cu) are 
correctly grouped together respectively in cluster 4, 1 and 3 
while the last two families are identified with clusters 2 and 5 with 
some difficulties.}
\label{fig:clusters}
\end{figure}

As in the work by \cite{Paccanaro}, we use the Astral 95 compendium of
SCOP database \cite{SCOP} where no two proteins share more than 95\%
of similarity, so as not to overload the clustering procedure with
huge numbers of very similar proteins that could easily be attributed
to a cluster by direct comparison if necessary. As this dataset is
hierarchically organised, we choose to work at the level of
superfamilies, in the sense that we want identify, on the basis of
sequence content, which proteins belong to the same
superfamily. Proteins belonging to the same superfamily are
evolutionary related and share functional properties. Before going
into the detail of the results we want to underline the fact that we
do not modify our algorithm to adapt to this dataset structure, and
without any prior assumption on the data, we are able to extract
interesting information on the relative size and number of clusters
selected (Fig.\ref{fig:clusters}).
Notably we do not use a training set to optimise a model of the
underlying cluster structure, but focus only on raw sequences and
alignments.

One issue that was recently put forward is the alignment variability
\cite{KarenM} depending on the algorithms employed. Indeed some of our
results could be biased by errors or dependence of the dissimilarity
matrix upon the particular details of the alignments that are used to
compute distances, but in the framework of a clustering procedure
these small-scale differences should stay unseen due to the large
scale of the dataset. On the other hand, the great advantage of
working only with sequences is the opportunity to use our method on
datasets where no structure is known a priori, such as fast developing
metagenomics datasets \cite{Venter:2004}.
\begin{figure}
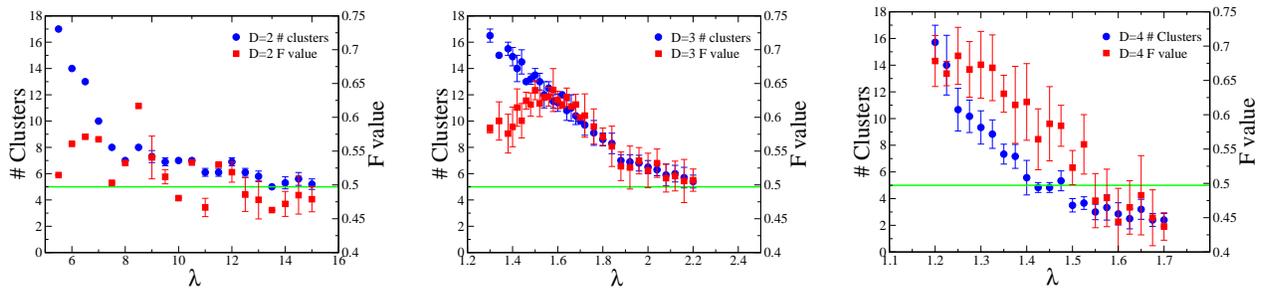

\epsfig{file=Figure/dis3.eps,scale=0.2}\hspace{0.5cm}
\epsfig{file=Figure/dis4.eps, scale=0.2}\hspace{0.5cm}
\epsfig{file=Figure/dis5.eps,scale=0.2}
\caption{In the three panels we show the average number of clusters
  over the random noise as a function of the weight of the root for
  $D=2,3,4$ respectively.  For each graph we show the number of
  clusters (circle) and the associated F value (square), computed as a
  function of precision and recall. We want to emphasise the fact the
  highest F values are reached for depth $D=4$ and weight $\lambda\sim
  1.3$.  With this choice of the parameters we found the number of
  clusters is of order $10$, a good approximation of the number
  of superfamilies shown in figure as a straight line.}
\label{num_clu}
\end{figure}
We choose as a training set $5$ different superfamilies belonging to
the ASTRAL 95 compendium for a total number of 661 proteins: a)
Globin-like, b) EF-hand, c) Cupredoxin, d) Trans-Glycosidases and e)
Thioredoxin-like. Our algorithm is able to identify a good
approximation on the real number of clusters. Here we choose the
parameter $\lambda$ well above the typical weight between different
nodes, so as to minimise the number of groups found. As a function of
this weight you can see the number of clusters found by the $D$-MST
algorithm reported in figure \ref{num_clu}, for the depths
$D=2,3,4$. In these three plots we see the real value of the number
of clusters is reached for different values of the weight $\lambda
\sim 12,2,1.4$ respectively. The performance of the algorithm can be
analysed in terms of precision and recall. These quantities are
combined in the $F$-value \cite{Paccanaro} defined as
\begin{equation}
F= \frac{1}{N}\sum_h n_h \max_i \frac{2 n_i^h}{n^h+n_i},
\end{equation}
where $n_i$ is the number of nodes in cluster $i$ according to the
classification $\lambda$ we find with the $D$-MST algorithm, $n^h$ is
the number of nodes in the cluster $h$ according to the real cluster
classification $K$ and $n_i^h$ is the number of predicted proteins in
the cluster $i$ and at the same time in the cluster $h$.  In both
cases the algorithm performs better results for lower value of
$\lambda$. This could be related to the definition of the $F$ value
because starting to reduce the number of expected clusters may be
misleading in the accuracy of the predicted data
clustering. 

 Since distances between datapoints have been normalised to be
real numbers between $0$ to $1$, when $\lambda\to\infty$ we expect to find the 
number of connected components of the given graph $G(n,s)$. While lowering
this value, we start to find some configurations which minimise the
weight respect to the single cluster solution. The role played by the
external parameter $\lambda$ could be seen as the one played by a
chemical potential tuning from outside the average number of clusters.

\begin{figure}
\begin{center}
\epsfig{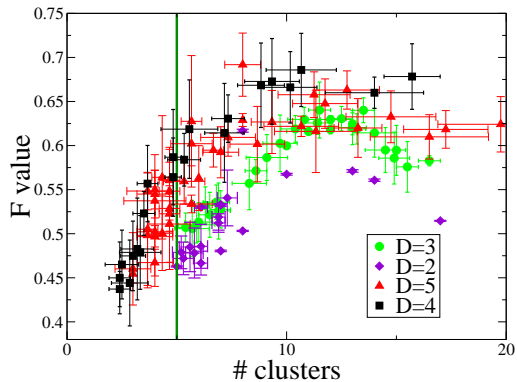}
\end{center}
\caption{We plot the F value for depths $D=2,3,4,5$ as a
  function of the number of clusters found by the $D$-MST
  algorithm. The case $D=2$ provides the AP results while $D>n$ is
  associated to SL and gives value well below $0.4$. The highest 
  performance in terms of the F value is reached for depth
  $D=4$ and number of clusters $\sim 10$. We draw a line in
  correspondence to the presumed number of clusters which is $5$ where again the
  algorithm with parameter $D=4$ obtains the highest performance
  score.}
\label{fig:fvalue}
\end{figure}
We compare our results to the ones in \cite{Paccanaro} for
different algorithms and it is clear that intermediate values of $D$ 
gives best results on the number of clusters detected and on the $F$-value
reached without any a priori treatment of data.  It is also clear that
$D$-MST algorith mwith $D=3,4,5$ gives better results than AP 
(case $D=2$) as can be seen in Fig. \ref{fig:fvalue}. 

We believe that the reason is that clusters do not have an intrinsic 
spherical regularity. This may be due to the fact that
two proteins having a high number of differences between their
sequences at irrelevant sites can be in the same family. Such
phenomena can create clusters with complex topologies in the sequence
space, hard to recover with methods based on a spherical shape
hypothesis.  We compute the $F$-value also in the single linkage limit
($D>n$) and its value is almost $\sim 0.38$ in all the range of
clusters detected. This shows that the quality of the predicted clusters
improves reaching the highest value when $D=4$ and then decreases
when the maximum depth increases.

\subsection{Clustering of verbal autopsy data}

The verbal autopsy is an important survey-based approach to measuring
cause-specific mortality rates in populations for which there is no
vital registration system \cite{MurrayLFPY2007, KingL2008}.  We
applied our clustering method to the results of 2039 questionnaires in
a benchmark verbal autopsy dataset, where gold-standard
cause-of-death diagnosis is known for each individual.  Each entry in
the dataset is composed of responses 47 {\it yes/no/don't know}
questions.

To reduce the effect of incomplete information, we restricted our analysis to the
responses for which at least 91\% of questions answered yes or no (in
other words, at most 9\% of the responses were ``don't know'').  This
leaves 743 responses to cluster (see \cite{MurrayLFPY2007} for a
detailed descriptive analysis of the response patterns in this
dataset.)

The goal of clustering verbal autopsy responses is to infer the common
causes of death on the basis of the answers.  This could be used in
the framework of ``active learning'', for example, to identify which
verbal autopsies require further investigation by medical professionals.

As in the previous applications, we define a distance matrix on the
verbal autopsy data and apply $D$-MST with different depths $D$.  The questionnaires are turned into vectors by associating  to the answers  yes/no/don't know the values 0/1/0.5 respectively.
The similarity matrix is  then computed  as the root mean square difference between vectors,
$ d_{ij}=\frac{1}{N}\sqrt{\sum_k \left(s_i(k)-s_j(k)\right)^2}$, where $s_i(k)\in\{0,1,0.5\}$ refers to the symptom $k \in [0,47]$ in the i-th questionnaire.

\begin{figure}
\begin{center}
    \epsfig{file=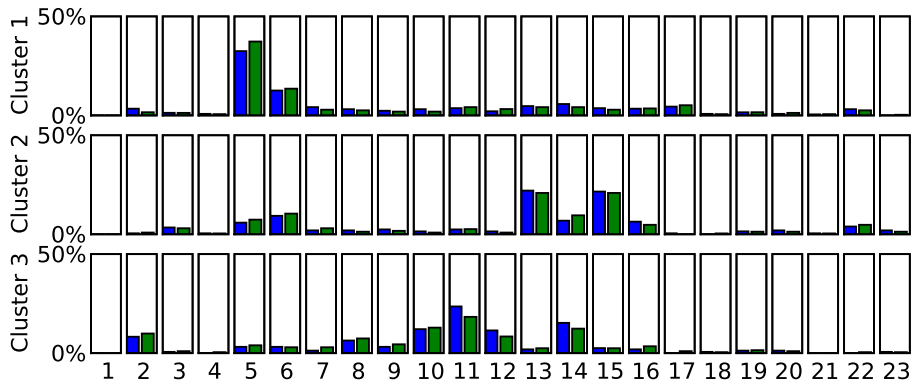,scale=0.7}
 \end{center}
  \caption{Cluster decomposition broken down by cause-of-death  (from 1 to 23) produced   by  AP (blue) and $D$-MST (green).  The parameter $\lambda$  is chosen from the stable region,  where the number  of clusters is constant.}
    \label{fig:dead}
\end{figure}

We first run $2$-MST (AP) and $4$-MST on the dataset and find how the number of clusters depend on $\lambda$.  We identify  a stable region which corresponds to 3 main clusters for  both $D=2,4$. 
As shown in figure \ref{fig:dead}, to each cluster we can associate a different causes of death. Cluster 1 contains nearly all of the 
Ischemic Heart Disease deaths (cause 5) and about half of the Diabetes Mellitus deaths (cause 6).  Cluster 2 contains most of the Lung Cancer
deaths (cause 13) and Chronic Obstructive Pulmonary Disease deaths (cause 15).  Cluster 2 also contains most of the additional IHD and DM
deaths (30\% of all deaths in the dataset are due to IHD and DM). Cluster 3 contains most of the Liver Cancer deaths (cause 11) as well
as most of the Tuberculosis deaths (cause 2) and some of the other prevalent causes.
For $D=2$ we find no distinguishable hierarchical structure in the 3 clusters, while for higher value we find a second-level structure. In particular for $D=4$ we obtain 57-60 subfamilies for value of $\lambda$ in the region of $0.15-0.20$.
Although the first-level analysis (Fig.\ref{fig:dead}B) underlines the  similarity of $D$-MST algorithm with AP, increasing the depth leads to a finer sub-clusters decomposition \cite{progress}.

\subsection{Conclusion}

We introduced  a new clustering algorithm which naturally interpolates between partitioning methods and hierarchical clustering. The algorithm is based on the cavity method and finds  a bounded  depth $D$ spanning tree on 
a graph $G(V,E)$ where $V$ is the set of $n$ vertices identified with  the data points plus one additional root node and  
$E$ is the set of edges with weights given by the dissimilarity matrix and by a unique distance $\lambda$ from the root node.
The limits $D=2$ and $D=n$ reduce to the well known AP and SL algorithms. The choice of  $\lambda$ determines the number of clusters. Here we have adopted  the same criterion as in ref. \cite{Frey2007}: the first non trivial clustering occurs when the cluster number is constant for a stable region of $\lambda$ values.

Preliminary applications on three different biological datasets have shown that  it is indeed possible to exploit the deviation from the purely $D=2$ spherical limit to gain some insight  into the data structures.
Our method has properties which are of generic relevance for large scale datasets,  namely  scalability, simplicity and parallelizability. Work is in progress to systematically apply this technique to real world data.

\begin{acknowledgments}
The work was supported by a Microsoft External Research Initiative grant.
SB acknowledges MIUR grant 2007JHLPEZ.
\end{acknowledgments}

\bibliographystyle{unsrt}
\bibliography{biblio}

\end{document}